\documentclass[aip,jap,amsmath,amssymb,reprint]{revtex4-1}
\usepackage{graphicx}
\usepackage{bm}
\usepackage{caption}
\usepackage{amsmath}
\usepackage[mathlines]{lineno}

\begin{document}
\title[Effect of Size and Shape on Electronic and Optical Properties of CdSe Quantum Dots]{Effect of Size and Shape on Electronic and Optical Properties of CdSe Quantum Dots}
\author{Yincheng Liu}
\email{liu\_yincheng@rvhs.edu.sg}
\affiliation{ 
	Nanyang Technological University, 50 Nanyang Avenue, Singapore 639798}
\affiliation{River Valley High School, 6 Boon Lay Avenue, Singapore 649961}
\author{Sumanta Bose}
\email{sumanta001@e.ntu.edu.sg}
\affiliation{ 
	Nanyang Technological University, 50 Nanyang Avenue, Singapore 639798}
\author{Weijun Fan}
\email{ewjfan@ntu.edu.sg}
\affiliation{ 
Nanyang Technological University, 50 Nanyang Avenue, Singapore 639798}

\date{\today}

\begin{abstract}
We used an effective mass envelope function theory based on the 8-band \textbf{\textit{k$\cdot$p}} model with valence force field considerations to investigate the effect of size and shape on electronic and optical properties of cadmium selenide quantum dots. Major factors related to their  properties including band mixing probabilities, spatial charge distributions, transition matrix elements and Fermi factors were studied. Volumetrically larger CdSe dots were found to have smaller band-gaps but higher transition matrix elements and Fermi factors. The maximum optical gain for dots was observed to have an initially positive and then negative correlation with their real-space size as a result of combined effects of various factors. For the shape effects, cubic dots were found to have smaller band-gaps, Fermi factors and transition matrix elements than spherical dots due to higher level of asymmetry and surface effects. Consequently, cubic dots have lower emission energy, smaller amplification but broader gain spectrum. Cubic and spherical dots are both promising candidates for optical devices under visible range. In this study, we have demonstrated  that size and shape change could both effectively alter the properties of quantum dots and therefore recommend consideration of both when optimizing the performance for any desired application.
\end{abstract}


\maketitle
\section{\label{sec:level1}Introduction}
Cadmium selenide (CdSe) quantum dots have gained vast amount of research attention over the recent years. As II-VI semiconductors, their innate large band-gap allows their great potential in optoelectronic industries, such as laser and light-emitting devices (LEDs) under the visible spectrum range\cite{RN35}. CdSe quantum dots have been found to possess significantly improved performance compared to conventional bulk semiconductor, including higher tunability\cite{RN4}, energetic efficiency\cite{RN5} and optical amplification\cite{RN40} due to their three-dimensional quantum confinement effects. Recent advancements in nanocrystal synthesis have demonstrated the feasibility of accurate size\cite{RN6} and shape\cite{RN41} control of CdSe quantum dots using colloidal synthesis. The size and shape changes were expected to profoundly affect the properties of CdSe dots. However, at the current stage, despite that the effect of size on quantum dots has been studied thoroughly\cite{RN42}, the research work done on the shape effect is still limited.

Optoelectronic properties of CdSe quantum dots heavily depend on their band structure and band-mixing probabilities, which are sensitive towards size and shape changes. A few methods have been developed to determine their electronic structures, including empirical pseudo-potential theory\cite{RN8}, effective mass approach\cite{RN9} and multi-band \textbf{\textit{k$\cdot$p}} methods\cite{RN36}. All three methods generally predict similar qualitative features. However, persisting minor difference exist for each energy level calculated for a particular quantum dot morphology. The \textbf{\textit{k$\cdot$p}} method is most popular as it automatically considers the piezoelectric effect of lattice strain\cite{RN10}. In this study, we used an effective mass envelope function theory approach based on the 8-band \textbf{\textit{k$\cdot$p}} method to compute the electronic properties of CdSe quantum dots. The subsequent computation of optical properties took into account of transition matrix element\cite{RN11}, carrier density\cite{RN12}, Fermi factor\cite{RN13} and homogeneous broadening\cite{RN15} which were determined to have profound influence over optical gain spectrum. The findings of this study could provide insights to use shape control during quantum dots synthesis as an alternative, instead of solely relying on the size change, to alter the properties of CdSe quantum dots for various industrial purposes.

\section{\label{sec:level2}Methodology}
\begin{figure}[h!]	
	\includegraphics[width=0.4\textwidth]{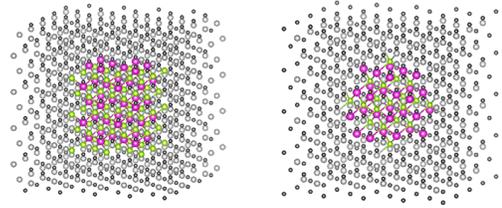}
\captionsetup{justification=centering}
\centering
 \caption{A 3D atom-by-atom model of cubic and spherical CdSe Quantum Dot.}
	\label{F1}
\end{figure}

CdSe quantum dots considered in this study are assumed to have the zincblende lattice structure and be colloidally synthesized. We studied the cubic quantum dots with length from 4 to 12 units of lattice constants (corresponding volume: 14-383 nm\(^3\)) and spherical dots of the same volume. The temperature is assumed to be 300K (room temperature), as electronic and optical properties of quantum dots are temperature-sensitive. Fig.\ref{F1} shows an atom-by-atom model of typical CdSe quantum dots under infinite potential well studied in this project. For each quantum dot geometry, we applied an effective mass envelope wavefunction theory approach based on 8-band \textbf{\textit{k$\cdot$p}} methods with the consideration of  orbit-splitting effects to obtain the bandstructure near the \(\Gamma\)-point of the Brillouin zone. The 8-band Hamiltonian is represented in the Bloch function basis of \(|S\rangle \uparrow, |P_x\rangle \uparrow,|P_y\rangle \uparrow,|P_z\rangle \uparrow,|S\rangle \downarrow, |P_x\rangle \downarrow,|P_y\rangle \downarrow,|P_z\rangle \downarrow \) as \cite{RN18}
\newline

\begin{widetext}
\begin{equation}
	{H_=\begin{pmatrix}
	h_{11} & \frac{ip_{0}(k'_{x}+ik'_{y})}{\sqrt{2}} & ip_{0}k'_{z} & \frac{ip_{0}(k'{x}-ik'_{y})}{\sqrt{2}} & 0 & 0 & 0 &0\\ \frac{ip_{0}(k'_{x}+ik'_{y})}{\sqrt{2}} & h_{22} & h_{23} & h_{24} & 0 & 0 & 0 & 0 \\ -ip_{0}k'_{z} & h_{23} & h_{33} & h_{34}& 0 & 0 & 0 & 0\\ \frac{ip_{0}(k'{x}-ik'_{y})}{\sqrt{2}} & h_{24} & h_{34} & h_{44} & 0 & 0 & 0 & 0\\ 0 & 0 & 0 & 0 & h_{55} & \frac{ip_{0}(k'_{x}+ik'_{y})}{\sqrt{2}} & ip_{0}k'_{z} & \frac{ip_{0}(k'{x}-ik'_{y})}{\sqrt{2}}\\  0 & 0 & 0 & 0 &\frac{ip_{0}(k'_{x}+ik'_{y})}{\sqrt{2}} & h_{66} & h_{67} & h_{68}\\0 & 0 & 0 & 0 & -ip_{0}k'_{z} & h_{76} & h_{77} & h_{78}\\0 & 0 & 0 & 0 &\frac{ip_{0}(k'{x}-ik'_{y})}{\sqrt{2}} & h_{86} & h_{87} & h_{88}
	\end{pmatrix}+H_{so}+V_0}
\end{equation}
\end{widetext}
where \(H_{so}\) is the Hamiltonian for orbit-splitting, \(E_p\) is the matrix element of Kane's theory and \(p_0=\sqrt{E_p/2m_e}\). Each element of the Hamiltonian Matrix is given below
\begin{multline}
h_{11}=h_{55}=E_{g}-\frac{\hbar^2}{2m_0}\gamma_c(k^2_x+k^2_y+k^2_z)+a_c[tr(\varepsilon)]
\end{multline}
\begin{multline}
h_{22}=h_{44}=h_{66}=h_{88}=-\frac{\hbar^2}{2m_{0}}\Big[\frac{L'+M'}{2}(k_{x}^2+k_{y}^2)\\+M'k_{z}^2]+a_{v}[tr(\varepsilon)]+\frac{b}{2}[tr(\varepsilon)-2\epsilon_{zz}\Big]
\end{multline}
\begin{multline}
		h_{23}=h_{34}=h_{67}=h_{78}=-\frac{\hbar^2}{2m_{0}}\Big[\frac{N'(k_{x}-ik_{y})}{\sqrt{2}}\Big] \\+\sqrt{6}d(\varepsilon_{xz}-i\varepsilon_{yz})
\end{multline}
\begin{multline}
	h_{24}=h_{86}=-\frac{\hbar^2}{2m_{0}}[\frac{L'-M'}{2}(k^{2}_{x}-k^{2}_{y})-iN'k_{x}k_{y}]\\+\frac{3b}{2}(\varepsilon_{xx}-\varepsilon_{yy})-i\sqrt{12}d\varepsilon_{xy}
\end{multline}
\begin{multline}
	h_{33}=h_{77}=-\frac{\hbar^2}{2m_{0}}[M'(k^{2}_{x}+k^{2}_{y}+L'k^{2}_{z})]+a_{v}[tr(\varepsilon)]\\+b[3\varepsilon_{zz}-tr(\varepsilon)]
\end{multline}
\begin{equation}
\gamma_{c}=\frac{m_{0}}{m^{*}_{e}}-\frac{E_{p}}{3}\Big[\frac{3E_{g}+2\Delta_{so}}{E_{g}(E_{g}+\Delta_{so})}\Big]
\end{equation}
\(k_x', k_y', k_z'\) are modified wavevectors, given by
\begin{equation}
k'_{x}=k_{x}-\epsilon_{xx}k_{x}-\epsilon_{xy}k_{y}-\epsilon_{xz}k_{z}
\end{equation}
\begin{equation}
k'_{y}=k_{y}-\epsilon_{xy}k_{x}-\epsilon_{yy}k_{y}-\epsilon_{yz}k_{z}
\end{equation}
\begin{equation}
k'_{z}=k_{z}-\epsilon_{xz}k_{x}-\epsilon_{yz}k_{y}-\epsilon_{zz}k_{z}
\end{equation} 
where \(\varepsilon\) is the strain matrix, \(a_c\) and \(a_v\) are the hydrostatic deformation potentials and \(b\) and \(d\) are the shear deformation potential. \(L', M'\) and \(N'\) are modified Luttinger parameters derived from effective mass parameters from  Lifshitz-Kosevich theory.

CdSe quantum dots are assumed to have three-dimensionally periodical arrangement and periodicities are given by \(L_x, L_y\) and \(L_z\). As such, the envelope wavefunction for quantum dots can be expressed as \cite{RN17}
\begin{equation}
\Psi_{m}=\{\Psi_{m}^{j}\}\{j=1,2...,8\}
\end{equation}
\begin{multline}
\Psi_{m}^{j}=\frac{1}{\sqrt{V}}\sum_{n_{x},n_{y},n_{z}}a_{m,n_{x},n_{y},n_{z}}^{j}exp\Big[i(k_{nx}x\\+k_{ny}y+k_{nz}z)\Big]
\end{multline}
where \(V=L_xL_yL_z\), \(k_{ni}=2\pi n_i/L_i\), \(n_i\) is the plane-wave number in the range of \(\pm 3\). \(j\) and \(m\) are indexes for energy subband and for basis. Direct-diagonalization was used to solve for the eigenvalue of the Hamiltonian matrix.  

We used the valence force field model to estimate strain relaxation of quantum dots. The total elastic energy is calculated as a sum of bond-bending and bond-stretching effects of each atom, expressed as \cite{RN19}
\begin{multline}
	E=\frac{1}{2}\sum_{i(j)} \frac{3a_{ij}}{8d_{0,ij}^2}\Big(|\mathbf{r_{i}-r_{j}}|^2-d_{0,ij}^2\Big)^2+\sum_{i(j,k)}\\\frac{3\beta_{jik}}{8d_{0,ij}d_{0,ik}}\Big(\mathbf{|r_{i}-r_{j}}||\mathbf{r_{i}-r_{k}}|+\frac{d_{0,ij}d_{0,ik}}{3}\Big)^2
\end{multline}
where \(i,j\) and \(k\) are indexes for adjacent atoms in CdSe crystal lattice, \(d_{0,ij}\) is the theoretical atomic bond length and \(\mathbf{r_i-r_k}\) is the displacement vector between 2 atoms.\(\alpha_{ij} \) and \(\beta_{ijk}\) are respectively the bond-stretching constant and bond-bending constant between atom \(i,j\) and \(k\).

The linear optical gain of CdSe quantum dots active region was computed based on density-matrix equation, with the consideration of homogeneous expansion of Lorentz shape, expressed as\cite{RN20}
 \begin{equation}
 G(E)=\frac{2\pi e^2 \hbar}{cn_{r} \varepsilon_{0} m_{e}^{2} \Omega} \sum_{c,v}\frac{|P_{cv}|^{2}(f_c-f_v)}{E_{cv}}B_{cv}(E-E_{cv})
 \label{gain}
 \end{equation}
 where \(n_r\) is the refractive index, \(\varepsilon_0\) is the dielectric constant, \(c\) is the speed of light, \(E_{cv}\) is the transition energy and \(\Omega\) is the real-space volume of quantum dot. \(f_c-f_v\) is the Fermi factor which gives out the rate of stimulated emission. \(f_c\) and \(f_v\) are probabilities of electrons occupying conduction and valence bands, determined by Fermi-Dirac distribution, \cite{RN37}
 \begin{equation}
 f_{i}=\frac{1}{1+exp[E_{ni}-E_{fi}/k_{B}T]}
 \label{fer1}
 \end{equation}
 where \(i=c,v\), \(E_{ni}\) is the quantized energy level and \(E_{fi}\) is quasi-Fermi level of of each discrete band. \(B_{cv}(E-E_{cv})\) accounts for the homogeneous broadening of transition energy, given by \cite{RN38}
 \begin{equation}
 B_{cv}(E-E_{cv})=\frac{ \Gamma_{cv}/2\pi}{(E-E_{cv})^2+( \Gamma_{cv}/2)^2}
 \end{equation}
 where the dephasing constant, \(\Gamma_{cv}\) is related with intraband relaxation time, \(\tau_r\), as \(\Gamma_{cv}=\hbar/\tau\). \(P_{cv}\) accounts for the transition matrix element, which describes the momentum of transition between each discreet electron and hole band. It is given by \cite{RN21}
 \begin{equation}
 P_{cv,i}=\langle \Psi_{c,k}|e_i \cdot p|\Psi_{v,k}\rangle, i=x,y,z
 \label{tme}
 \end{equation}
 Average of \(P_{cv,x}\) and \(P_{cv,y}\) accounts for the gain in the transverse electric mode and \(P_{cv,z}\) accounts for the gain in transverse magnetic mode.
 \begin{figure}[t!]	
 	\centering
 	\includegraphics[width=0.45\textwidth]{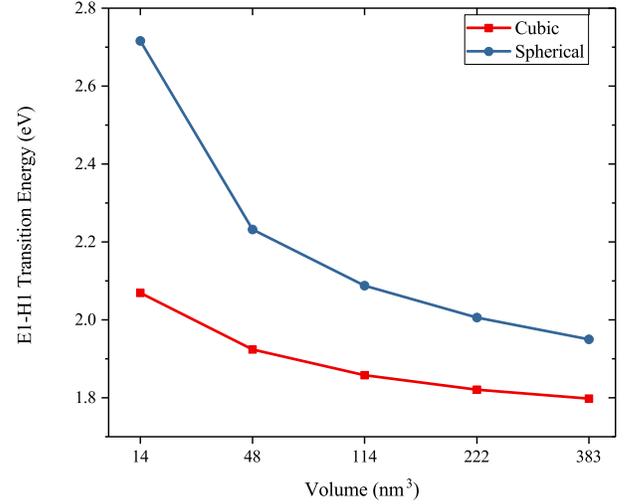}
 	\captionsetup{justification=centering}
 	\caption{E1-H1 Transition Energy for Cubic and Spherical CdSe Quantum Dots}
 	\label{F2}
 \end{figure}

\section{\label{sec:level3}Results and Discussion}

\subsection{\label{sec:level4}Electronic Properties}
\begin{figure}[t!]	
	\centering
    \includegraphics[width=0.45\textwidth]{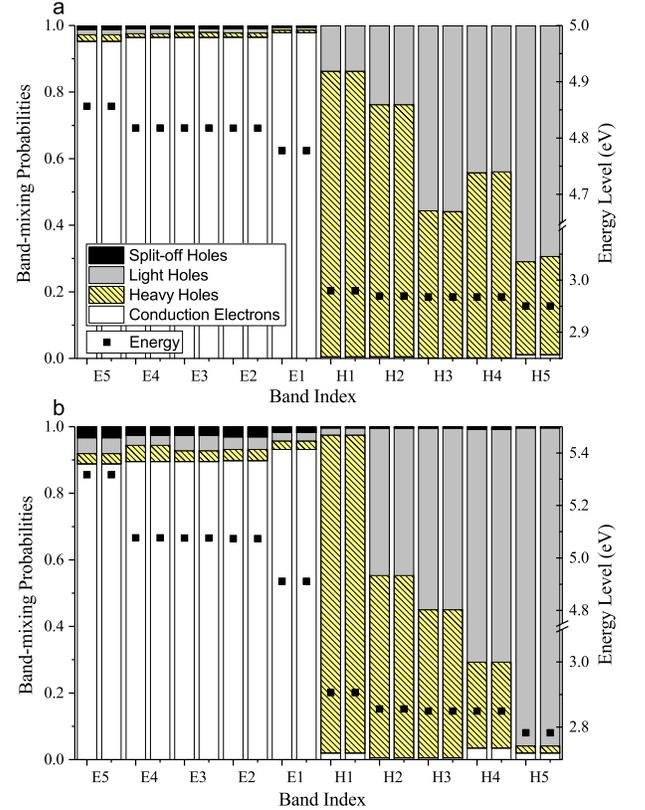}
	\captionsetup{justification=centering}
	\caption{Band Structure and Band-mixing Probabilities for Cubic (a) and Spherical (b) Quantum Dots of volume \(221.7\) nm\(^3\)}
	\label{F3}
\end{figure}
We varied the volumetric size of cubic and spherical quantum dots to investigate their effects on QD electronic properties. Fig.\ref{F2} shows the transition energy between lowest conduction band (E1) and highest valence band (H1) as a function of the real-space volume of quantum dots. E1-H1 transition energy is of great significance as it dictates the wavelength of primary transition. The plot shows that the transition energy has a negative correlation with the volumetric size of quantum dots for each geometry, as expected and seen from  previous work. However, a clear distinction between transition energy of the cubic and spherical dots of the same real-space volume can be observed, as spherical dots are found to have significantly higher values. This is because cubic dots have higher degree of asymmetry that weakens their quantum confinement effects.\cite{RN22,RN25} Furthermore, higher surface-to-volume ratio of spherical dots will make them more susceptible towards surface effects which also explains their larger band-gaps. \cite{RN39}

Fig.\ref{F3} shows the band structure and band-mixing probabilities of split-off holes, light holes, heavy holes and conduction electrons of bottom 5 conduction bands (E1-E5) and top 5 valence bands (H1-H5) for cubic and spherical quantum dots with the same real-space volume of  \(221.7\) nm\(^3\). Band-mixing probabilities depend on the extent of coupling of electrons and holes in each discreet band. Similar patterns of band-mixing probabilities between cubic and spherical dots of different sizes can be observed due to the absence of lattice-mismatch induced strain. The dominance of heavy holes in more energetically accessible top valence bands indicates that transverse electric mode (\(x-y\) polarized) gain will be dominant for each geometry\cite{RN27}. Smaller and spherical dots have larger intra-subband energy gaps due to their stronger quantum confinement effects. It is found that the intra-subband energy gaps between conduction bands are much larger than those between the valence bands. This is due to the larger effective mass of heavy holes than light holes that results in flatter dispersion curve of the conduction bands.

\begin{figure}[t!]	
	\centering
	\includegraphics[width=0.45\textwidth]{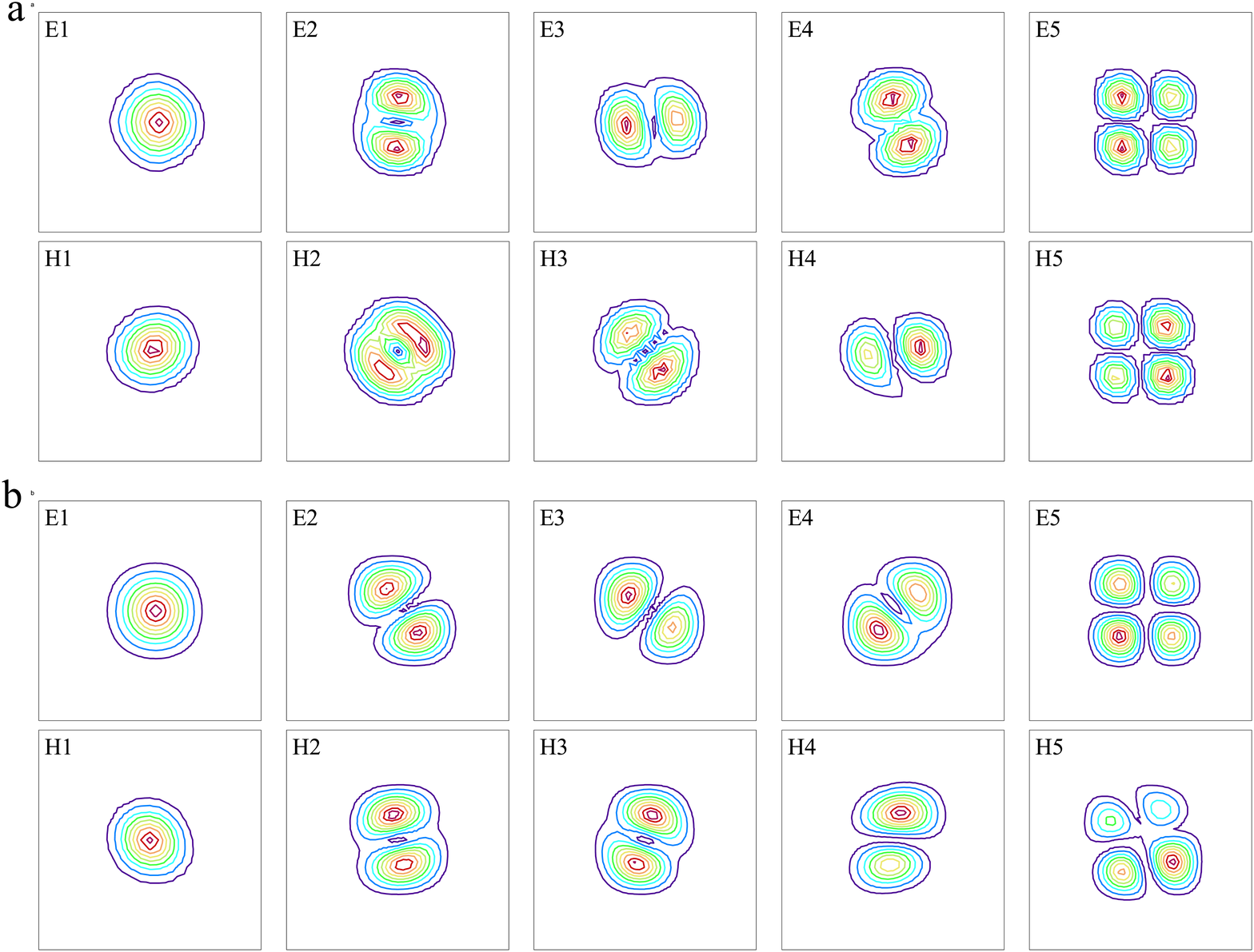}
	\captionsetup{justification=centering}
	\caption{Spatial Charge Density for (a) cubic and (b) spherical CdSe Quantum Dots of volume \(221.7\) nm\(^3\).}
	\label{F4}
\end{figure}

Fig.\ref{F4} shows the spatial charge density of bottom 5 conduction bands (E1-E5) and top 5 valence bands (H1-H5) of cubic and spherical quantum dots with the same real-space volume of  \(221.7\) nm\(^3\). Spatial charge density was calculated by squaring the wavefunction. It is of interest as the effectiveness of hole-electron wavefunction overlap determines the transition matrix elements and therefore the strength of optical emission. Bottom conduction bands (E1) and top valence bands (H1) are found to be \(s\)-like while bands at excited states are different extents of mixing between \(s, p\) and \(d\)-like distribution. Stronger mixing can be observed from cubic dots because of their relatively smaller intra-subband energy gap. Strongest transitions of cubic and spherical quantum dots are both E1-H1 transition in TE mode while the strongest TM mode transition is expected to be E1-H\(_n\), which is expected to be\(s\)-like and case dependent.

\begin{figure}[t!]	
	\centering
	\includegraphics[width=0.45\textwidth]{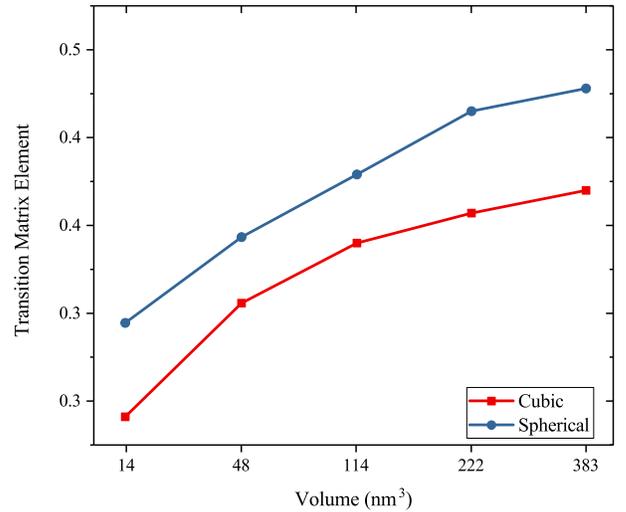}
	\captionsetup{justification=centering}
	\caption{Transition Matrix Element of E1-H1 Transition for Cubic and Spherical CdSe Quantum Dots}
	\label{F5}
\end{figure}

The transition matrix element of E1-H1 transition for each geometry was calculated using Eq.\ref{tme} and plotted in Fig.\ref{F5}. Increase in volumetric size for both cubic and spherical dots results in a clear increase in matrix elements as the wavefunction overlap becomes more extensive for volumetrically larger dots. The increase becomes gradual for larger dots as the value approaches unity. Spherical dots are found to have significantly higher matrix elements than cubic dots. This is because cubic dots have higher level of asymmetry that weakens wavefunction overlap. It is also reported by Andreev (2005) that sharp edges of cubic dots would result in piezoelectric effects that further explains the lower E1-H1 transition matrix elements of cubic dots. \cite{RN11}
\begin{figure*}[t!] 
	\includegraphics[width=1.0\textwidth]{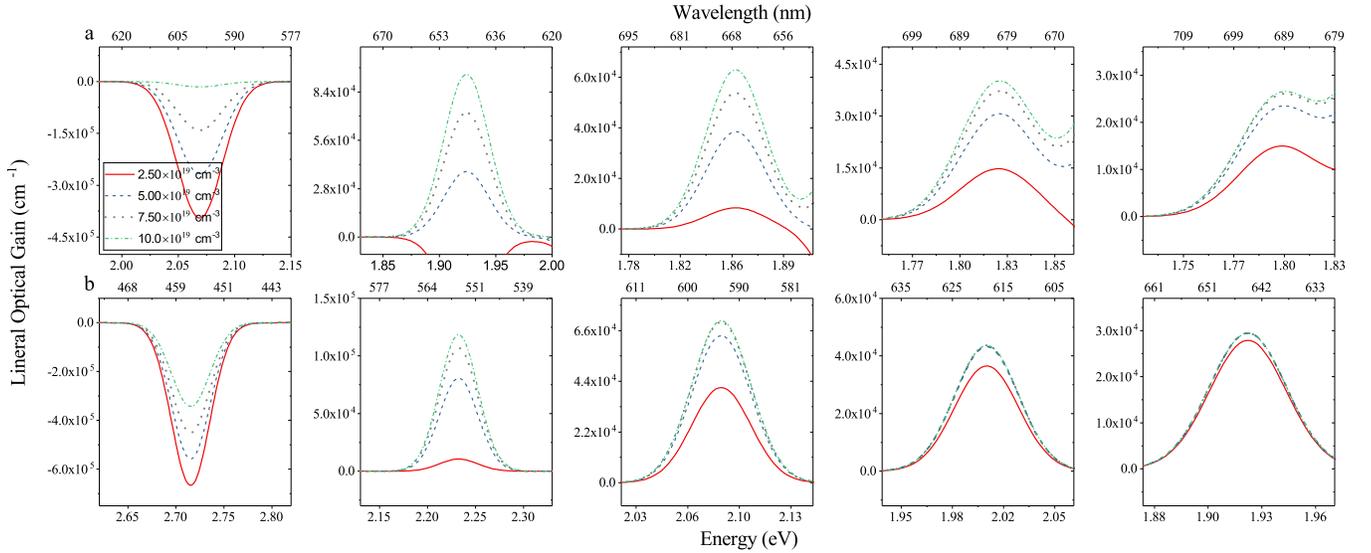}
	\caption{Optical Gain Spectrum for Cubic (a) and Spherical (b) Quantum Dots of Volume 14.2 nm\(^3\), 47.8 nm\(^3\), 113.5 nm\(^3\), 221.7 nm\(^3\) and 383.0 nm\(^3\) (from left to right) }
	\label{F6}
\end{figure*}
\subsection{Optical Properties}
CdSe quantum dots have demonstrated their great potential in functioning as optical devices, such as optical fiber amplifier, low threshold laser and LED. In this paper, we investigated the effect of size and shape change on the optical properties of CdSe quantum dots. The optical gain spectrum for differently sized cubic and spherical quantum dots was plotted in Fig.\ref{F6}. Carrier density for each case was varied to investigate the relationship between optical gain and lasing threshold current density.

Refer to Eq.\ref{gain}, the linear optical gain is dependent on the real-space volumes of dots, band-gaps, transition matrix elements and Fermi factors. Band-gap size dictates the peak position of the gain spectrum. It was previously determined that band-gap has a negative correlation with the size due to lower quantum confinement effects in Fig.\ref{F2}. Further, spherical dots were found to have larger band-gaps. The observations of Fig.\ref{F6} correspond to the aforementioned points that red-shift of optical gain spectrum is shown as volumetric size increases and emission from spherical dots are of larger energy than cubic dots. The extent of red-shift is significant smaller for cubic dots as \(\Delta E_{14.2 nm^3-383.0nm^3}\) is 271 meV for cubic dots, which is much smaller than 791 meV for spherical dots. This shows that the emission spectrum of cubic dots tends to be less sensitive towards the size change.

\begin{figure}[t!]
	\includegraphics[width=0.45\textwidth]{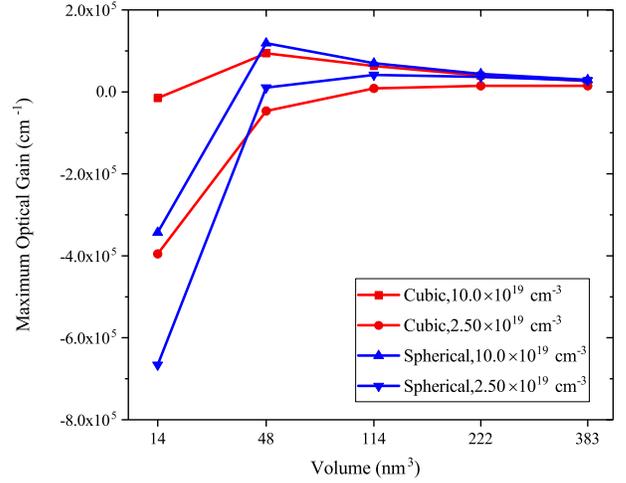}
	\caption{Maximum Optical Gain for Cubic and Spherical CdSe Quantum Dots}
	\label{F7}
\end{figure}

\begin{figure}[t!]
	\includegraphics[width=0.45\textwidth]{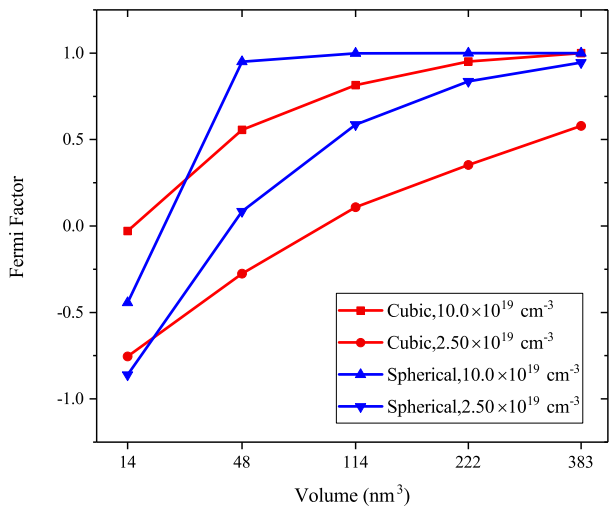}
	\caption{Fermi Factor for Cubic and Spherical CdSe Quantum Dots}
	\label{F8}
\end{figure}

Maximum gain of each quantum dot is determined by the combination of these factors. From Fig.\ref{F7}, extremely small dots (14.3 nm\(^3\)) display absorption as number of injected carriers is insufficient to overcome their wide band-gaps. The initial increase in volume leads to higher maximum gain for both cubic and spherical dots as larger dots have higher transition matrix elements (shown in Fig.\ref{F5}) and Fermi factors. Fig.\ref{F8} shows that Fermi factor has a positive correlation with the volume as conduction band becomes more energetically accessible due to smaller band-gaps. Spherical dots are found to have larger Fermi factor in spite of their larger band-gaps because their wider intra-subband energy gaps facilitates the population inversion between E1 and H1 bands. Maximum gain decreases as the size increases beyond 113.5 nm\(^3\). By this stage, both Fermi factor and transition matrix element approach saturation and increase in volume becomes the dominant factor. Dots larger than 113.5 nm\(^3\) can therefore be considered as less QD-like. Spherical dots are found to have higher maximum gain due to their larger Fermi factor and transition matrix element.

The effect of carrier density on optical gain was also considered. As shown in Fig.\ref{F8}, higher carrier density leads to higher Fermi factor and thus higher maximum gain. Gain saturation can also be observed in volumetrically larger dots that further increase in carrier density would not result in higher maximum gain due to the full occupancy of E1 and H1 bands. We notice the band-filling effects as carrier density increases for any quantum dot that a slight blue-shift in peak position is observed for higher carrier density. This is because electrons' occupancy of higher electronic states under high carrier density increases the effective energy gap between LUMO and HOMO.\cite{MELLITI200522} This phenomenon is however less prominent for larger quantum dots due to saturation. It is noteworthy that the cubic dots with volume of 383.0 nm\(^3\) shows the occurrence of near-E1-H1 transition which effectively increases the width of gain spectrum. Such phenomenon was not observed in smaller and spherical dots because their excited states transition possesses much higher energy and is further away from primary transition.

\begin{figure}[t!]
	\includegraphics[width=0.465\textwidth]{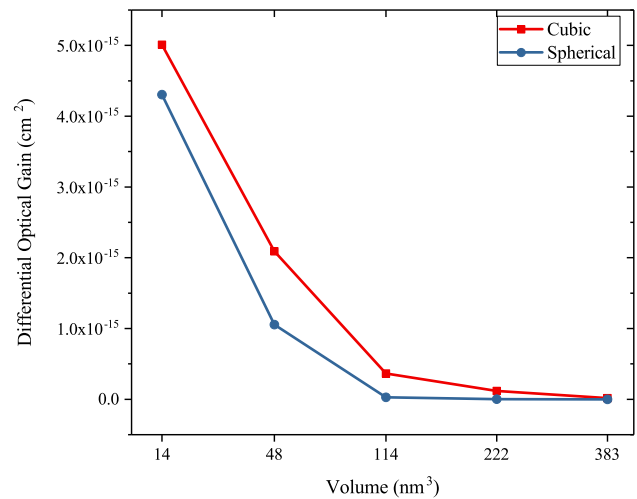}
	\caption{Differential Optical Gain for Cubic and Spherical CdSe Quantum Dots}
	\label{F9}
\end{figure}

The efficiency of quantum dots was determined with their differential optical gain and transparency carrier density. Differential optical gain was calculated as a derivative of maximum gain with respect to carrier density. Higher differential gain leads to high modulation bandwidth, lower frequency chirp and narrower line-width capabilities\cite{last}. Fig.\ref{F9} presents the differential gain for cubic and spherical dots at carrier density of 10.0\(\times 10^{19}\) cm\(^{-3}\). Volumetrically larger quantum dots are found to have lower value. Similarly, values for spherical dots are lower than those of cubic dots. Both observations can be explained by higher level of saturation in these quantum dots. Corresponding values for lower carrier densities were also calculated and similar pattern was observed.

\begin{figure}[t!]
	\includegraphics[width=0.48\textwidth]{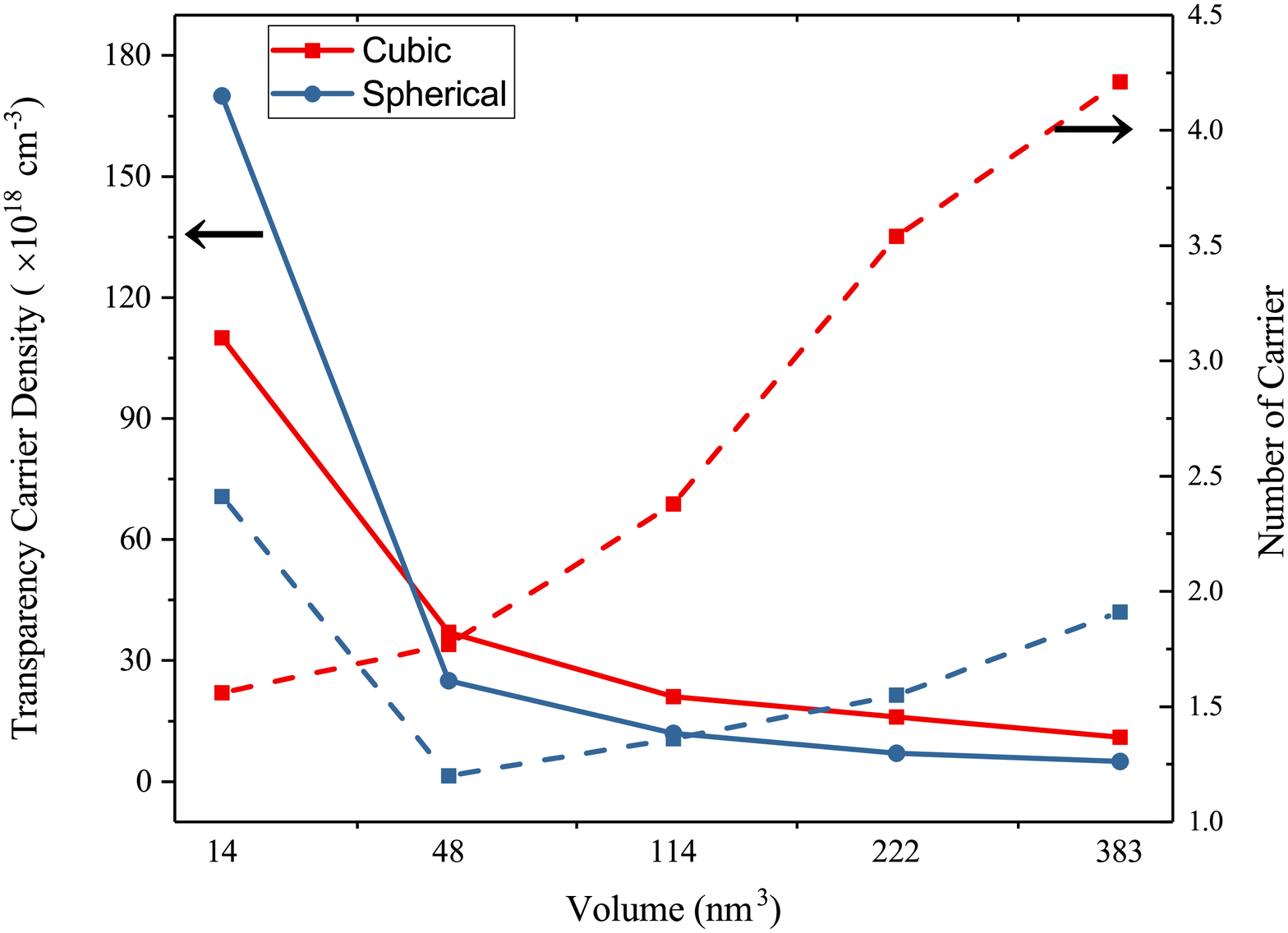}
	\caption{Transparency Carrier Density (left axis) and Corresponding Carrier Number (right axis) for Cubic and Spherical CdSe Quantum Dots}
	\label{F10}
\end{figure}

Transparency carrier density for quantum dots is the minimum carrier density required to reach the transparency condition, \(V_t=\frac{E_g}{q}\), when there is sufficient exciton count to obtain transparency.\cite{trans} Transparency carrier density for each QD geometry is affected by both band-gap and intra-subband energy gap. In this study, transparency carrier density was determined by extrapolating maximum optical gain against their respective carrier density.  In Fig.\ref{F10}, we plotted the calculate transparency carrier density (solid line) and corresponding carrier number (dashed line) for each CdSe quantum dot. In general, transparency carrier density has a negative correlation with the size due to decrease in quantum confinement effects. Generally, spherical dots have lower transparency carrier densities due to their larger intra-subband energy gap, which facilitates the occurrence of population inversion. Carrier number was calculated as the product of transparency density and real-space volume of the quantum dot. We notice that number of carriers required for transparency increases with volumetric size due to loss in quantum efficiency. Similarly, spherical dots are found to require smaller number of carriers to reach transparency due to their larger intra-subband energy gap, indicating their better energetic efficiency than cubic dots.

\section{Conclusion}
We performed a comprehensive and detailed study on the effect of size and shape change on the electronic properties, including transition energy, band mixing probabilities, spatial charge density and on the optical properties, including optical gain spectrum, Fermi factor, differential gain, transparency condition, of CdSe quantum dots. 

Data obtained in Section \ref{sec:level3} has shown that variation in size and shape leads to profound effects on the properties of quantum dots. Larger dots have higher transition matrix element and Fermi factor while the trade-off is lower efficiency. Similarly, spherical dots are found to have higher optical amplification while cubic dots have wider gain spectrum. Both cubic and spherical dots are promising candidates for high-performance optical devices under the visible light spectrum and more superior than conventional bulk semiconductor in terms of optical amplification and energetic efficiency. However, spherical dots are expected to have better performance because of their higher transition matrix element and Fermi factor. 

Furthermore, results obtained could provide insights for industry to use shape control as an alternative method to optimize the performance for any desired application, which is often neglected in current practices. This study has demonstrated that shape change could avoid the trade-off between emission energy and gain intensity. Furthermore, shape control can be effective in varying electronic properties and optical properties of quantum dots. Therefore, it is recommended to use the combination of size and shape control in order to achieve desired performance.

\section{Appendixes}

Constants related to material parameters of CdSe used for calculation in Section \ref{sec:level2} were presented in Table I. We used modified Luttinger Parameters, \(L', M'\) and \(N'\) for our 8-band \textbf{\(k\cdot p\)} model, which are derived from \(\gamma_1, \gamma_2\) and \(\gamma_3\) as
\begin{equation}
L'=L-\frac{E_{p}}{E_{g}}=-\frac{\hbar^{2}(\gamma_{1}+4\gamma_{2}+1)}{2m_{0}}
\end{equation}
\begin{equation}
M'=-\frac{\hbar^{2}(\gamma_{1}-2\gamma_{2}+1)}{2m_{0}}
\end{equation}
\begin{equation}
N'=-\frac{3\hbar^{2}\gamma_{3}}{m_{0}}
\end{equation}
\begin{table}[h]
	\caption{\label{tab:table1}Constants Used for Computation }
	\begin{ruledtabular}
		\begin{tabular}{lcr}
			Parameter & Symbol(Unit) & Value  \\\hline
			Lattice Constant & \(a_0(\AA)\)  & 6.052 \\  

			Effective Electron Mass & \(m_{e}^{*}(m_e)\) & 0.12 \\
		
			Kane Matrix Element & \(E_{p}(eV)\) & 16.5\\
		
			Refractive Index & \(n_r\) & 2.5\\
		
			Bulk Band Gap & \(E_g(meV)\)& 1732\\ 
			Spin-orbit Splitting Energy & \(\Delta_{so}(meV)\) & 420\\\hline
			Luttinger Parameter & \(\gamma_1\) & 3.265\\
			& \(\gamma_2\)& 1.162\\
			& \(\gamma_3\) & 1.443\\
			\hline 
			Deformation Potential & \(a_{c}(meV)\) & -2832.4\\
			& \(a_{v}(meV)\) & 1148.7\\
			& \(b(meV)\) & -1047.6\\ & \(d(meV)\) & -3100.0\\
			\hline 
			Valence Force Field Parameters & \(d_{Cd-Se}($\AA$)\) & 2.62\\
			& \(a_{Cd-Se}\) & 35.22\\
			& \(\beta_{Cd-Se-Cd}\) & 3.14\\
		\end{tabular}
	\end{ruledtabular}

\end{table}

\pagebreak

\section*{References}

\bibliographystyle{ieeetr}
\bibliography{bib2.bib}

\begin{thebibliography}{10}

\bibitem{RN35}
K.~Surana, P.~K. Singh, H.-W. Rhee, and B.~Bhattacharya, ``Synthesis,
  characterization and application of cdse quantum dots,'' {\em Journal of
  Industrial and Engineering Chemistry}, vol.~20, no.~6, pp.~4188--4193, 2014.

\bibitem{RN4}
E.~Y. Ko, J.~I. Lee, J.-W. Jeon, I.~H. Lee, Y.~H. Shin, and I.~K. Han, ``Size
  tunability and optical properties of {CdSe} quantum dots for various growth
  conditions,'' {\em Journal of the Korean Physical Society}, vol.~62, no.~1,
  pp.~121--126, 2013.

\bibitem{RN5}
H.~Lee, M.~Wang, P.~Chen, D.~R. Gamelin, S.~M. Zakeeruddin, M.~Grätzel, and
  M.~K. Nazeeruddin, ``Efficient {CdSe} quantum dot-sensitized solar cells
  prepared by an improved successive ionic layer adsorption and reaction
  process,'' {\em Nano Letters}, vol.~9, no.~12, pp.~4221--4227, 2009.

\bibitem{RN40}
C.~Wang, B.~L. Wehrenberg, C.~Y. Woo, and P.~Guyot-Sionnest, ``Light emission
  and amplification in charged cdse quantum dots,'' {\em The Journal of
  Physical Chemistry B}, vol.~108, no.~26, pp.~9027--9031, 2004.

\bibitem{RN6}
G.~Zlateva, Z.~Zhelev, R.~Bakalova, and I.~Kanno, ``Precise size control and
  synchronized synthesis of six colors of {CdSe} quantum dots in a
  slow-increasing temperature gradient,'' {\em Inorganic Chemistry}, vol.~46,
  no.~16, pp.~6212--6214, 2007.

\bibitem{RN41}
J.~G. Keizer, M.~Bozkurt, J.~Bocquel, T.~Mano, T.~Noda, K.~Sakoda, E.~C. Clark,
  M.~Bichler, G.~Abstreiter, J.~J. Finley, W.~Lu, T.~Rohel, H.~Folliot,
  N.~Bertru, and P.~M. Koenraad, ``Shape control of quantum dots studied by
  cross-sectional scanning tunneling microscopy,'' {\em Journal of Applied
  Physics}, vol.~109, no.~10, p.~102413, 2011.

\bibitem{RN42}
S.~Neeleshwar, C.~L. Chen, C.~B. Tsai, Y.~Y. Chen, C.~C. Chen, S.~G. Shyu, and
  M.~S. Seehra, ``Size-dependent properties of {CdSe} quantum dots,'' {\em
  Phys. Rev. B}, vol.~71, p.~201307, May 2005.

\bibitem{RN8}
G.~Bester, S.~Nair, and A.~Zunger, ``{Pseudopotential calculation of the
  excitonic fine structure of million-atom self-assembled
  In$_{1-x}$Ga$_x$As/GaAs quantum dots},'' {\em Physical Review B}, vol.~67,
  no.~16, p.~161306, 2003.

\bibitem{RN9}
M.~A. Cusack, P.~R. Briddon, and M.~Jaros, ``{Electronic structure of InAs/GaAs
  self-assembled quantum dots},'' {\em Physical Review B}, vol.~54, no.~4,
  pp.~R2300--R2303, 1996.

\bibitem{RN36}
M.~W. Lok C. Lew Yan~Voon, {\em The k p Method-Electronic Properties of
  Semiconductors}.
\newblock Springer-Verlag Berlin Heidelberg, 2009.

\bibitem{RN10}
M.~Ehrhardt, {\em Multi-Band Effective Mass Approximations}.
\newblock Lecture Notes in Computational Science and Engineering, Springer
  International, 2014.

\bibitem{RN11}
A.~D. Andreev, ``{Optical matrix element in InAs∕GaAs quantum dots:
  Dependence on quantum dot parameters},'' {\em Applied Physics Letters},
  vol.~87, no.~21, p.~213106, 2005.

\bibitem{RN12}
M.~Röwe, P.~Michler, J.~Gutowski, V.~Kümmler, A.~Lell, and V.~Härle,
  ``Influence of the carrier density on the optical gain and refractive index
  change in ingan laser structures,'' {\em physica status solidi (a)},
  vol.~200, no.~1, pp.~135--138, 2003.

\bibitem{RN13}
S.~Nakamura and S.~F. Chichibu, {\em Introduction to nitride semiconductor blue
  lasers and light emitting diodes}.
\newblock London ; New York: Taylor \& Francis, 2000.

\bibitem{RN15}
T.~Liu, K.~E. Lee, and Q.~J. Wang, ``Microscopic density matrix model for
  optical gain of terahertz quantum cascade lasers: Many-body, nonparabolicity,
  and resonant tunneling effects,'' {\em Physical Review B}, vol.~86, no.~23,
  p.~235306, 2012.

\bibitem{RN18}
S.~Bose, Z.~Song, W.~J. Fan, and D.~H. Zhang, ``Effect of lateral size and
  thickness on the electronic structure and optical properties of quasi
  two-dimensional cdse and cds nanoplatelets,'' {\em Journal of Applied
  Physics}, vol.~119, no.~14, p.~143107, 2016.

\bibitem{RN17}
J.~Chen, W.~J. Fan, Q.~Xu, X.~W. Zhang, and S.~S. Li, ``{Electronic structure
  and optical gain saturation of InAs$_{1−x}$N$_x$/GaAs quantum dots},'' {\em
  Journal of Applied Physics}, vol.~105, no.~12, p.~123705, 2009.

\bibitem{RN19}
H.~Jiang and J.~Singh, ``{Strain distribution and electronic spectra of
  InAs/GaAs self-assembled dots: An eight-band study},'' {\em Physical Review
  B}, vol.~56, no.~8, pp.~4696--4701, 1997.

\bibitem{RN20}
M.~Sugawara, K.~Mukai, Y.~Nakata, H.~Ishikawa, and A.~Sakamoto, ``{Effect of
  homogeneous broadening of optical gain on lasing spectra in self-assembled
  In$_x$Ga$_{1-x}$As/GaAs quantum dot lasers},'' {\em Physical Review B},
  vol.~61, no.~11, pp.~7595--7603, 2000.

\bibitem{RN37}
Y.~Shapira and L.~J. Neuringer, ``{Experimental Verification of the Fermi-Dirac
  Distribution of Electrons in Gallium},'' {\em Phys. Rev. Lett.}, vol.~18,
  pp.~1133--1135, Jun 1967.

\bibitem{RN38}
S.~Koirala, S.~Mouri, Y.~Miyauchi, and K.~Matsuda, ``Homogeneous linewidth
  broadening and exciton dephasing mechanism in
  $\mathrm{MoT}{\mathrm{e}}_{2}$,'' {\em Phys. Rev. B}, vol.~93, p.~075411, Feb
  2016.

\bibitem{RN21}
M.~Rohlfing and S.~G. Louie, ``Electron-hole excitations and optical spectra
  from first principles,'' {\em Physical Review B}, vol.~62, no.~8,
  pp.~4927--4944, 2000.

\bibitem{RN22}
L.~Liang and W.~Xie, ``Influence of the shape of quantum dots on their optical
  absorptions,'' {\em Physica B: Condensed Matter}, vol.~462, pp.~15--17, 2015.

\bibitem{RN25}
L.~H. Peng, Y.~C. Hsu, and C.~W. Chuang, ``Structure asymmetry effects in the
  optical gain of piezostrained ingan quantum wells,'' {\em IEEE Journal of
  Selected Topics in Quantum Electronics}, vol.~5, no.~3, pp.~756--764, 1999.

\bibitem{RN39}
H.~F. Wilson, L.~McKenzie-Sell, and A.~S. Barnard, ``Shape dependence of the
  band gaps in luminescent silicon quantum dots,'' {\em J. Mater. Chem. C},
  vol.~2, pp.~9451--9456, 2014.

\bibitem{RN27}
P.~Zhao, L.~Han, M.~R. McGoogan, and H.~Zhao, ``{Analysis of TM mode light
  extraction efficiency enhancement for deep ultraviolet AlGaN quantum wells
  light-emitting diodes with III-nitride micro-domes},'' {\em Optical Materials
  Express}, vol.~2, no.~10, pp.~1397--1406, 2012.

\bibitem{MELLITI200522}
A.~Melliti, M.~Maaref, B.~Sermage, J.~Bloch, F.~Saidi, F.~Hassen, and
  H.~Maaref, ``{Thermal emission and band-filling effects on the
  photoluminescence rise time of InGaAs/InAs/GaAs quantum dots},'' {\em Physica
  E: Low-dimensional Systems and Nanostructures}, vol.~28, no.~1, pp.~22 -- 27,
  2005.

\bibitem{last}
S.~Selmic, {\em Multiple Quantum Well Semiconductor Lasers Emitting at
  Wavelengths of 1310 Nm and 1550 Nm}.
\newblock 2002.

\bibitem{trans}
O.~Svelto, {\em Introductory Concepts}.
\newblock Boston, MA: Springer US, 2010.

\end{thebibliography}

\end{document}